\documentclass[12pt,a4paper]{article}
\begin{document}
\textwidth=135mm
 \textheight=200mm
\begin{center}
{\bfseries Bogoliubov compensation approach in QCD and in the electro-weak theory
\footnote{{\small Talk at the International Bogolyubov Conference "Problems of Theoretical and Mathematical Physics", Moscow-Dubna, Russia, August 21-27 2009.}}}
\vskip 5mm
B.A. Arbuzov$^{\dag}$ and M.
K. Volkov$^\ddag$
\vskip 5mm
{\small {\it $^\dag$ Skobeltsyn Institute of Nuclear Physics of MSU, 119991 Moscow, Russia}}\\
{\small {\it $^\ddag$ Joint Institute for
Nuclear Research, 141980 Dubna, Russia}} \\
\end{center}
\vskip 10mm

In previous works~\cite{Arb04, Namb2, Arvol, Arbplb, Arvol2, Arbepj}
N.N. Bogoliubov compensation principle~\cite{Bog1, Bog2}
was applied to studies of spontaneous generation of effective non-local interactions in renormalizable gauge theories.
In view
of this one performs "add and subtract" procedure for the effective
interaction with a form-factor. Then one assumes the presence of the
effective interaction in the interaction Lagrangian and the same term with
the opposite sign is assigned to the newly defined free Lagrangian. 

Now we shall firstly demonstrate an application of the principle to  the electro-weak theory.
We start with EW Lagrangian with $N_{gen}=3$ lepton and colour quark doublets
with gauge group $SU(2)$. That is we restrict the gauge sector to
triplet of $W^a_\mu$ only.
\begin{eqnarray}
& & L\,=\,\sum_{k=1}^3\biggl(\frac{\imath}{2}
\Bigl(\bar\psi_k\gamma_
\mu\partial_\mu\psi_k\,-\partial_\mu\bar\psi_k
\gamma_\mu\psi
_k\,\biggr)-\,m_k\bar\psi_k\psi_k\,+
\,\frac{g}{2}\,\bar\psi_k\gamma_\mu \tau^a W_\mu^a\psi_k
\biggr)\,+\nonumber\\
& & +\sum_{k=1}^3\biggl(\frac{\imath}{2}
\Bigl(\bar q_k\gamma_
\mu\partial_\mu q_k\,-\partial_\mu\bar q_k
\gamma_\mu q_k\,\biggr)-\,M_k\bar q_k q_k\,+
\,\frac{g}{2}\,\bar q_k\gamma_\mu \tau^a W_\mu^a q_k
\biggr)\,-\nonumber\\
& &-\frac{1}{4}\,\biggl( W_{\mu\nu}^aW_{\mu\nu}^a\biggr);\qquad
W_{\mu\nu}^a\,=\,
\partial_\mu W_\nu^a - \partial_\nu W_\mu^a\,+g\,\epsilon_{abc}W_\mu^b W_\nu^c\,.
\label{initial}
\end{eqnarray}
where we use the standard notations and $\psi_k$ and $q_k$ correspond to left leptons and quarks respectfully. We write here masses for leptons and quarks bearing in mind the ready Higgs phenomenology.
In accordance to the Bogoliubov approach in application to
QFT we look for
a non-trivial solution of a
compensation equation, which is formulated on the basis
of the Bogoliubov procedure add -- subtract. We have $L\,=\,L_0\,+\,L_{int}$, where
\begin{eqnarray}
& &L_0\,=\,=\,\sum_{k=1}^3\biggl(\frac{\imath}{2}
\Bigl(\bar\psi_k\gamma_
\mu\partial_\mu\psi_k\,-\partial_\mu\bar\psi_k
\gamma_\mu\psi
_k\,\biggr)-\,m_k\bar\psi_k\psi_k\,+\frac{\imath}{2}
\Bigl(\bar q_k\gamma_
\mu\partial_\mu q_k\,-\nonumber\\
& &\partial_\mu\bar q_k
\gamma_\mu q_k\,\biggr)-\,M_k\bar q_k q_k\biggr)\,-\,\frac{1}{4}\,W_{\mu\nu}^a W_{\mu\nu}^a\,+
\,\frac{G}{3!}\cdot\,\epsilon_{abc}\,W_{\mu\nu}^a\,W_{\nu\rho}^b\,W_{\rho\mu}^c\,;
\label{L0}\\
& &L_{int} = \frac{g}{2} \sum_{k=1}^3\biggl(\bar\psi_k\gamma_\mu
\tau^a W_\mu^a\psi_k + \bar q_k\gamma_\mu \tau^a W_\mu^a q_k\biggr) - \frac{G}{3!}\cdot \epsilon_{abc} 
W_{\mu\nu}^a W_{\nu\rho}^b W_{\rho\mu}^c\,.\label{Lint}
\end{eqnarray}
Here notation
 $\frac{G}{3!}\cdot \,\epsilon_{abc}\,
W_{\mu\nu}^a\,W_{\nu\rho}^b\,W_{\rho\mu}^c$ means corresponding
non-local vertex in the momentum space
\begin{eqnarray}
& &(2\pi)^4 G\,\epsilon_{abc} (g_{\mu\nu} (q_\rho pk - p_\rho qk)+ g_{\nu\rho}
(k_\mu pq - q_\mu pk)+g_{\rho\mu} (p_\nu qk - k_\nu pq)+\nonumber\\
& &+q_\mu k_\nu p_\rho - k_\mu p_\nu q_\rho)\,F(p,q,k)\,
\delta(p+q+k)\,+...\;;\label{vertex}
\end{eqnarray}
where $F(p,q,k)$ is a form-factor and
$p,\mu, a;\;q,\nu, b;\;k,\rho, c$ are respectfully incoming momenta,
Lorentz indices and weak isotopic indices
of $W$-bosons.

Effective interaction $-\,\frac{G}{3!}\cdot \,\epsilon_{abc}\,
W_{\mu\nu}^a\,W_{\nu\rho}^b\,W_{\rho\mu}^c\,;\quad G\,=\,
\frac{g\,\lambda}{M_W^2}\,$ is
usually called  anomalous three-boson interaction and it is
considered for long time on phenomenological grounds.

Let us consider  expression~
(\ref{L0}) as the new  free Lagrangian $L_0$,
whereas expression~(\ref{Lint}) as the new
interaction Lagrangian $L_{int}$. 
Then compensation conditions will
consist in demand of full connected tree-boson vertices of the structure~(\ref{vertex}),
following from Lagrangian $L_0$, to be zero. This demand
gives a non-linear equation for form-factor $F$.

Such equations according to terminology by Bogoliubov 
 are called compensation equations.
In a study of these equations it is always evident the
existence of a perturbative trivial solution (in our case
$G = 0$), but, in general, a non-perturbative
non-trivial solution may also exist.

The goal of a study is a quest of an adequate
approach, the first non-perturbative approximation of
which describes the main features of the problem.
Improvement of a precision of results is to be achieved
by corrections to the initial first approximation.

Now in  first approximation we come the following
equation for $F(x)$
\begin{eqnarray}
& &F(z)\,=\,1 + \frac{85\, g\,\sqrt{N} \,\sqrt{z}}{96\,\pi}\Biggl(
\ln\,z + 4\,
\gamma + 4\,\ln\,2 +\frac{1}{2}\,G_{15}^{31}\Bigl( z_0\,|^0_{0, 0, 1/2, -1, -1/2}\Bigr) - \nonumber\\
& &-\,\frac{3160}{357}\Biggr) +
\frac{2}{3\,z} \int_0^z F(t)\,t\, dt - \frac{4}{3\,\sqrt{z}} \int_0^z F(t)
\sqrt{t}\, dt - \frac{4\,\sqrt{z}}{3} \int_z^{z_0} F(t) \frac{dt}{\sqrt{t}}\,+
\nonumber\\
& &+\,\frac{2\,z}{3}\,\int_z^{z_0}\,F(t)\,\frac{dt}{t}\,\quad
z\,=\,\frac{G^2\,N\,x^2}{1024\,\pi^2}\,.\label{eqFg}\\
& &\nonumber
\end{eqnarray}
Here $x = p^2$, $N=2$. We introduce here
an effective cut-off $Y$, which bounds a "low-momentum" region where
our non-perturbative effects act
and consider the equation at interval $[0,\, Y]$ under condition $F(Y)\,=\,0\,$.
We solve equation~(\ref{eqFg}) and obtain
\begin{eqnarray}
& &F(z)\,=\,\frac{1}{2}\,G_{15}^{31}\Bigl( z\,|^0_{1,\,1/2,\,0,\,-1/2,\,-1}
\Bigr) -\,\frac{85\,g \sqrt{N}}{512\,\pi}\,G_{15}^{31}\Bigl( z\,|^{1/2}_{1,\,1/2,
\,1/2,\,-1/2,\,-1}\Bigr)\,+\nonumber\\
& &+\,C_1\,G_{04}^{10}\Bigl( z\,|_{1/2,\,1,\,-1/2,\,-1}\Bigr)\,+
\,C_2\,G_{04}^{10}\Bigl( z\,|_{1,\,1/2,\,-1/2,\,-1}\Bigr)\,.
\label{solutiong}
\end{eqnarray}
where
$
G_{qp}^{nm}\Bigl( z\,|^{a_1,..., a_q}_{b_1,..., b_p}\Bigr)
$
is a Meijer function~\cite{be}. Constants $C_1,\,C_2$ are defined by boundary conditions. With $N\,=\,2$ this gives
\begin{equation}
g(z_0)\,=\,-\,0.4301\,;\; z_0\,=\,205.4254\,;\;
C_1\,=\,0.00369\,; \; C_2\,=\,0.00582\,.\label{gY}
\end{equation}
Note that there is also solution with large positive $g(z_0)$, which will be considered further on.

We use Schwinger-Dyson equation for $W$-boson polarization operator to
obtain a contribution of additional effective vertex to the running EW
coupling constant $\alpha_{ew}$. The contribution under
discussion reads after angular integrations
\begin{eqnarray}
& &\Delta \Pi_{\mu \nu}(x)\,=\,(g_{\mu \nu}\,p^2 -
p_\mu p_\nu)\,\Pi(x)\,;\quad x\,=\,p^2\,;\quad y'\,=\,q^2+\frac{3 x}{4}\,;
\nonumber\\
& &\Pi(x)\,=\,-\, \frac{g\,G\,N}{32\,\pi^2}\Biggl(
\int_{3 x/4}^x
\frac{F(y') dy'}{y'-x/2}\,\biggl(16\frac{y'^3}{x^2}-
48\frac{y'^2}{x}+45 y-
\frac{27}{2}x\biggr)\,+\nonumber\\
& &+\,\int_x^Y \frac{F(y') dy'}{y'-x/2}\,\biggl(-\,3 y'\,+\,\frac{5}{2}\,x
\biggr)\Biggr)\,; \quad g\,=\,g(Y)\,. \label{DF}
\end{eqnarray}
So we have modified one-loop expression for $\alpha_{ew}(p^2),\,x=p^2$
\begin{equation}
\alpha_{ew}(x)\,=\,\frac{6\,\pi\,\alpha_{ew}(x'_0)}{6\,\pi\,+\,5\,\alpha_{ew}(x_0')
\ln(x/x_0')\,+\,6\,\pi\,\Pi(x)}\,; \quad \alpha_{ew}(x_0)\,=\,\frac{g(Y)^2}{4\,\pi}\,; \label{al1}
\end{equation}
where $x_0'= 4/3\,x_0$ means a normalization point such that $\Pi(x'_0)=0$. 
Using expression~(\ref{al1}) we calculate behaviour of
$\alpha_{ew}(x)$ down to values of $x=M_W^2$ and obtain $\alpha_{ew}(M_W^2)\,=\,0.0374;\; (\alpha_{ew}^{exp}(M_W^2)\,=\,\frac{\alpha(M_W)}{\sin^2_W}\,=\,0.0337)\,$. It is only $10\%$ larger than the experimental value. We consider this result as strong confirmation of the approach.

Let us consider a contribution of effective interaction
in~(\ref{Lint}) to $g-2$ anomaly $\Delta a$. Considering contributions of the the new three-boson interaction to $W\,W\,H$ interaction we have from its conventional definition 
\begin{equation}
G\,=\,\frac{g \lambda}{M_W^2}\,;\quad \lambda\,=\,-\,0.0151\,;\label{lambda2}
\end{equation}
that agrees with experimental limitations.
Then for 
mass of Higgs $M_H\,=\,114\,GeV$ we obtain
$\Delta a\,=\,3.34\cdot 10^{-9}$,
that comfortably fits into error bars for well-known deviation
~\cite{g22}
$\Delta a\,=\,(3.02\,\pm\,0.88)\cdot 10^{-9}$.
With  $M_H$ growing $\Delta a$ slowly decreases  inside the error bars down to
$2.67\cdot 10^{-9}$ for $M_H = 300\,GeV$~\cite{Arbepj}.

The same procedure we apply to QCD Lagrangian with $N_f=3$ colour quarks with gauge group $SU(3)$ and study a possibility of a spontaneous generation of effective interaction
$-\,G_g/3!\cdot \,f_{abc}\,
F_{\mu\nu}^a\,F_{\nu\rho}^b\,F_{\rho\mu}^c$, which
 is usually called  anomalous three-gluon interaction.
The second solution of Eq.(\ref{eqFg}) for $N = 3$ reads
\begin{equation}
z_0\,=\,0.01784\,;\; g(z_0)\,=\,2.92145\,;\;
C_1\,=\,-\,18.8241\,;\; C_2\,=\,56.2171\,.\label{agY}
\end{equation}
Then we again use Schwinger-Dyson equation for gluon polarization operator to
obtain a contribution of additional effective vertex to the running strong 
coupling constant $\alpha_s$. 

So we have modified one-loop expression for $\alpha_s(p^2)$
\begin{equation}
\alpha_s(x)\,=\,\frac{4\,\pi\,\alpha_s(x'_0)}{4\,\pi\,+\,9\,\alpha_s(x_0')
\ln(x/x_0')\,+\,4\,\pi\,\Pi(x)}\,; \quad x=p^2\,; \label{aal1}
\end{equation}
where $x_0'$ means a normalization point such that $\Pi(x'_0)=0$. We normalize the running coupling
by conditions
\begin{equation}
\alpha_s(x_0)\,=\,\frac{g(Y)^2}{4\,\pi}\,=\,0.679185;\quad\alpha_s(x'_0)\,=\,\frac{4 \alpha_s(x_0) \pi(1+\Pi(x_0))}{4\pi+9\,\alpha_s(x_0)\ln(4/3)}\,.\label{analpha}
\end{equation}
Applying the standard transformation we have
\begin{equation}
\alpha_s(x)\,=\,\frac{4 \pi }{9\,\ln\frac{x}{\Lambda^2}\biggl(1+2\,g\,\sqrt{3}\,\biggl(\alpha_s(x'_0)\, \ln\frac{x}{\Lambda^2}\biggr)^{-1} \Pi(x)\biggr)}\,;\;\Pi(x) = 0\; for\;x\ge x'_0\, .\label{aall}\\
\end{equation}
From here we have also $\frac{x_0}{\Lambda^2}\,=\,1.1707002\,; \;G_g\,=\,\frac{6.62198}{\Lambda^2}$.
Using expressions~(\ref{aal1}, \ref{analpha}, \ref{aall}) and normalization at point $M_\tau$, we calculate behaviour of
$\alpha_s(x)$. We also use 
Shirkov-Solovtsov~\cite{SS} method to eliminate the ghost pole, that means the following substitution in (\ref{aall})
\begin{eqnarray}
& &\alpha_s(x)\,=\,\frac{4 \pi }{9}\biggl(\frac{1}{\ln\frac{x}{\Lambda^2}}\,-\,\frac{\Lambda^2}{x - \Lambda^2} \biggr)\biggl(1+\frac{2\,g\,\sqrt{3}}{\alpha_s(x'_0)}\,\biggl(\frac{1}{\ln\frac{x}{\Lambda^2}}\,-\,\frac{\Lambda^2}{x - \Lambda^2} \biggr) \Pi(x)\biggr)^{-1}\,;\nonumber\\
& &\alpha^{(2)}_s \simeq 0.67\; for\; Q\,<\,250\,MeV\, . \label{asubst}
\end{eqnarray}
  
As the next step we apply N.N. Bogoliubov compensation principle
to studies of spontaneous generation of effective non-local Nambu -- Jona-Lasinio interaction.
NJL model~\cite{NIL, VE} proves to be effective in description of 
low-energy hadron physics. 
However, the problem how to calculate parameters of the model ($G_i\,,\; \Lambda_i$\,;\,...) from the fundamental QCD was 
not solved for a long time. For the purpose 
we start with the conventional QCD Lagrangian with two light quarks and use again Bogoliubov procedure {\it add --
subtract} with test interaction
\begin{eqnarray}
& &\frac{G_1}{2}\cdot\Bigl(
\bar\psi\tau^b\gamma_5\psi\,\bar\psi \tau^b\gamma_5
\psi\,-\,\bar\psi\,\psi\,\bar\psi\,\psi\biggr)\,
+\,\nonumber\\
& &\frac{G_2}
{2}\cdot\Bigl(\bar\psi\tau^b\gamma_\mu\psi\,\bar\psi
\tau^b\gamma_\mu\psi +
\bar\psi \tau^b\gamma_5 \gamma_\mu \psi \bar\psi \tau^b
\gamma_5 \gamma_\mu \psi\biggr)\,;\label{addsub}
\end{eqnarray}
Here notation {\it e.g.}
 $\frac{G_1}{2}\cdot \,\bar\psi\,\psi\,\bar\psi\,\psi$ means corresponding
non-local vertex in the momentum space with form-factor 
$F_1(p_i)$

Then we again come to the corresponding compensation equation (see~\cite{Arvol}), which has the following non-trivial solution
\begin{eqnarray}
& &F_1(z)\,=\,C_1\,G^{40}_{06}\biggl( z|1,\frac{1}{2}, \frac{1}{2},0,a,b\biggr)+ C_2\,G^{40}_{06}\biggl( z|1,\frac{1}{2},b,a, \frac{1}{2},0\biggr)+\nonumber\\
& &C_3\,G^{40}_{06}\biggl( z|1,0,b,a,\frac{1}{2},\frac{1}{2}\biggr)\,;\; a=-\frac{1-\sqrt{1-64u_0}}{4};\; F_1(u_0)=1;\label{F1}\\
& &b=-\frac{1+\sqrt{1-64u_0}}{4};\; \beta = \frac{(G_1^2+6 G_1 G_2)N}{16 \pi^4};\;z=\frac{\beta x^2}{64};\;u_0=\frac{\beta m_0^4}{64}.\nonumber
\end{eqnarray}
Constants $C_i$ are defined by boundary conditions.
So we have the unique
solution. Value of parameter $u_0$ and ratio of two constants $G_i$
are also fixed $ u_0\,=\,1.92\cdot
10^{-8}\,\simeq 2\cdot 10^{-8}\,;\quad
G_1\,=\,\frac{6}{13}\,G_2\,.$ 
We would draw attention to a natural appearance 
of small quantity $u_0$.

Thus we come to effective non-local NJL interaction, which we use to obtain the description of low-energy hadron physics
~\cite{Namb2, Arvol, Arvol2}. 
In this way we obtain expressions for all quantities under study. Analysis shows that the optimal set of low-energy parameters  corresponds to $\alpha_s = 0.67$ and $m_0 = 20.3\,MeV$. We present a
set of calculated parameters for these conditions including
quark condensate, constituent quark mass $m$, parameters of light mesons 
\begin{eqnarray}
& &\alpha_s\,=\,0.67\,;\quad m_0\,=\,20.3\,MeV\,;
\quad m_\pi\,=\,135\,MeV\,;\quad g\,=\,3.16\,;  
\nonumber\\
& &m_\sigma\,=\,492\,MeV\,;
\quad \Gamma_\sigma\,=\,574\,MeV\,;\quad f_\pi\,=\,93\,
MeV\,;\nonumber\\
& &m\,=\,295\,MeV\,;\quad <\bar q\,q> =-\,(222\,MeV)^3\,;\quad G_1^{-1}\,=\,(244\,MeV)^2\,;\qquad.\nonumber\\
& &M_\rho\,=\,926.3\,MeV ;\quad \Gamma_\rho\,=\,159.5\,MeV \quad M_{a_1}\,=\,1174.8\,MeV ;\nonumber\\
& &\Gamma_{a_1}\,=\,350\,MeV ;\quad 
\Gamma(a_1\to \sigma\,\pi)/\Gamma_{a_1}\,=\,0.23\,.
\label{res}
\end{eqnarray}
We use input quantity $m_0$, while all other quantities are
calculated. The overall accuracy may be estimated to be of order of 10 -- 15\%. The worse accuracy 
occurs in value $M_\rho$ (20\%). 

\end{document}